\documentstyle[sprocl]{article}

\makeatletter
\newdimen\normalarrayskip              
\newdimen\minarrayskip                 
\normalarrayskip\baselineskip
\minarrayskip\jot
\newif\ifold             \oldtrue            \def\new{\oldfalse}
\def\arraymode{\ifold\relax\else\displaystyle\fi} 
\def\eqnumphantom{\phantom{(\theequation)}}     
\def\@arrayskip{\ifold\baselineskip\z@\lineskip\z@
     \else
     \baselineskip\minarrayskip\lineskip2\minarrayskip\fi}
\def\@arrayclassz{\ifcase \@lastchclass \@acolampacol \or
\@ampacol \or \or \or \@addamp \or
   \@acolampacol \or \@firstampfalse \@acol \fi
\edef\@preamble{\@preamble
  \ifcase \@chnum
     \hfil$\relax\arraymode\@sharp$\hfil
     \or $\relax\arraymode\@sharp$\hfil
     \or \hfil$\relax\arraymode\@sharp$\fi}}
\def\@array[#1]#2{\setbox\@arstrutbox=\hbox{\vrule
     height\arraystretch \ht\strutbox
     depth\arraystretch \dp\strutbox
     width\z@}\@mkpream{#2}\edef\@preamble{\halign
\noexpand\@halignto
\bgroup \tabskip\z@ \@arstrut \@preamble \tabskip\z@ \cr}%
\let\@startpbox\@@startpbox \let\@endpbox\@@endpbox
  \if #1t\vtop \else \if#1b\vbox \else \vcenter \fi\fi
  \bgroup \let\par\relax
  \let\@sharp##\let\protect\relax
  \@arrayskip\@preamble}
\def\eqnarray{\stepcounter{equation}%
              \let\@currentlabel=\theequation
              \global\@eqnswtrue
              \global\@eqcnt\z@
              \tabskip\@centering
              \let\\=\@eqncr
              $$%
 \halign to \displaywidth\bgroup
    \eqnumphantom\@eqnsel\hskip\@centering
    $\displaystyle \tabskip\z@ {##}$%
    \global\@eqcnt\@ne \hskip 2\arraycolsep
         $\displaystyle\arraymode{##}$\hfil
    \global\@eqcnt\tw@ \hskip 2\arraycolsep
         $\displaystyle\tabskip\z@{##}$\hfil
         \tabskip\@centering
    &{##}\tabskip\z@\cr}

\def\bea{\begin{eqnarray}}
\def\eea{\end{eqnarray}}
\def\nn{\nonumber}

\def\beq{\begin{equation}}
\def\eeq{\end{equation}}
\def\ba{\beq\new\begin{array}{c}}
\def\ea{\end{array}\eeq}
\def\be{\ba}
\def\ee{\ea}

\def\Tr{{\rm Tr}}
\def\res{{\rm res}}
\def\rank{{\rm rank}}
\def\2{{1\over 2}}

\begin{document}


\setcounter{footnote}0
\begin{center}
\hfill FIAN/TD-13/96,\ ITEP/TH-25/96\\
\vspace{0.3in}
{\LARGE\bf Integrability as Effective Principle of Nonperturbative
Field and String Theories
\footnote{based on a talk given at the Second International Sakharov
Conference 20-24 May 1996, Moscow, Russia}}\\
\bigskip
{\Large A.Marshakov}
\footnote{E-mail address:
mars@lpi.ac.ru}\\
{\it Theory Department,  P. N. Lebedev Physics
Institute , Moscow,~117924, Russia}\\
and {\it ITEP, Moscow 117259, Russia}\\
\end{center}
\bigskip


\begin{abstract}
One of the perspectives in modern quantum field and
string theory is related with the attempts to go beyond the perturbation
theory.
It turns out that a key principle in the formulation of all known
non-perturbative results is {\it integrability}, i.e.
arising of the structures of completely integrable systems. I discuss
several important steps in this direction and speculate on its
further possible development.
\end{abstract}





Recent investigations in gauge and string theories showed that sometimes
there exists a way to find explicitely the exact nonperturbative results
(spectrum, correlation functions, effective actions) even in the quantum
theories which can not be considered as {\it quantum integrable} models
at least in conventional and naive sense. In contrast to "naive"
quantum integrable models where usually the quantum (infinite-dimensional)
symmetry algebra allows one to compute explicitely the spectrum
and the correlation functions, there is no even to such extent "straightforward"
way in the theories I am going to discuss. The intention to study these
particular models is caused by the hope that they are not so far away from
the realistic quantum gauge and string theories where the solution to the
basic problems of confinement and quantum gravity is looked for, being
on the other hand {\it solvable} at least in the sense to be discussed below.

The starting, and maybe the most pessimistic point is that already in this
class of models
there are still no {\it direct} ways of solution -- like there are no such
ways for in many other
models of quantum field and string theory. However, this particular class
is distinguished by the fact (indeed a
strong hypothesis) that (generally some pieces of) the exact solutions do
really exist. A key reason for that is that the topological
excitations play an important role and determine the structure of the
theory at very low distances when the other excitations cancel each other
due to, say, (extended) supersymmetry.
Amazingly enough, all existing examples subjected to this scheme of
behaiviour have almost identically the {\it same effective} formulation
which can be presented in terms of effective completely integrable model
forgeting about many properties of the bare theory -- in particular there
is no real distinction between the theories living in diferent space-time
dimensions. The low-energy limit of such models is described by a sort
of {\it topological} theory.

Below, I will not pay enough attention to the mechanism which allows us to
distinguish this class of theories -- just mentioning again that $N \geq 2$
supersymmetry (which is always possible to find for such models at least as a
BRST-symmetry)
plays an important role. Instead I will try to describe the basic features
of the effective formulation, which attract a lot of interest themselves.

{\bf Moduli of the theory}. Any exact nonperturbative solution should present
the spectrum, correlation functions, effective actions as functions of the
parameters of the theory -- or its moduli, coming usually from the
(gauge-invariant) low-energy values of the fields in the target-space.
For example, in $4d$ supersymmetric gauge theories this is the v.e.v.'s
of the Higgs fields $h_k = {1\over k}\langle\Tr\Phi ^k\rangle$, in string
theories these are the moduli of the target-space metric (e.g. K\"ahler or
complex structures), gauge fields (e.g. moduli of flat connections or
selfdual gauge fields) etc. The problem
itself is to find the (exact nonperturbative) dependence of the physical
objects on these moduli parameters
\footnote{Of course, in addition to the parametric dependence of moduli
themselves the physical quantities can depend on the topological (discrete)
characteristics of moduli spaces, moreover in the simplest topological string
models only this dependence is essential and the correlation functions can be
just {\it numbers}.}.

{\bf Spectral curve}. Next point is that the class of the theories we discuss
is distinguished by the {\it holomorphic} dependence of the (complex)
moduli parameters -- i.e. there exists a complex structure on the moduli space
and only functions with "good" global behaiviour enter the game. This goes back to
the holomorphic structures arising in the instantonic calculus and to the
Belavin-Knizhnik theorem in string theory which leads to the draustical
simplifications removing the possible ambiguity in the form of the exact
answer. In the known examples, this holomorphic structures arise moreover
in a nice geometric way -- the moduli of the theory appear to be (a subspace
in) the moduli space of complex structures of the {\it target-\-space} spectral
curves $\Sigma $. It turns out, that {\it nonperturbatively} the target-\-space
spectral curve acquires a nontrivial topological structure (being just defined
locally -- as a
sphere of the scale parameter in perturbation theory) and a complex
structure on the spectral curve is parameterized by moduli of the theory. This
additional complicated structure
means that "stringy" nature plays an important role in the nonperturbative
formulation and strings (and $D$-branes) wraping along topologically
nontrivial directions
produce important effects in the exact effective formulation while
perturbatively the spectral curve can be tested only "locally".

{\bf Curves and Integrable Systems}.
Now let us discuss the {\it fact} of appearence of integrable systems
of KP/Toda type
in the framework of the quantum field and string theories.
Indeed, since the solutions are formulated in terms of periods of some
differentials on a complex spectral curve -- it means that an integrable system
(moreover the particular class of systems of KP/Toda type -- where the
Liouville torus is restricted to be a real section of a Jacobian of a complex
curve)
appears more or less by definition by means of the Krichever construction.
This very {\it useful} observation leads us to a
possibility of applying rather simple technique of Lax pairs, spectral
curves, symplectic forms and $\tau $-functions to naively infinite-dimensional
quantum field and string theories.

{\bf Symplectic geometry of integrable systems}.
To go further we will start with the point -- what should be added to a
complex curve to have an integrable system. Indeed, the Liouville torus as
a real section of a Jacobian is determined after one introduces following
Dubrovin, Krichever and Novikov a {\it meromorphic} 1-form $dS$ whose
derivatives ${\partial dS\over\partial h_k}
\cong \omega _k$  give {\it holomorphic} differentials. This generating 1-form
defines a completely integrable system on a symplectic manifold
$\Omega = \delta dS $ which in all cases we discuss can be explicitely
rewritten as
\be\label{dS}
dS = \lambda d\log w = \Tr{\cal L}d\log T
\nn \\
\Omega = \delta\lambda\wedge\delta\log w = \Tr\delta{\cal L}\wedge\delta\log T
\ee

{\bf Symplectic form and duality}.
The symplectic form (\ref{dS}) is defined by the eigenvalues of {\it two}
operators playing the essential role in integrable systems. Quasiclassically
their common spectrum defines the spectral curve. The
symplectomorphisms of (\ref{dS}) can be
considered as transformations between the {\it dual} integrable systems
with the generating function $S = \sum _k\int ^{\gamma _k}dS$.

Indeed, writing (\ref{dS}) more explicitely one finds that $\Omega = \sum _k
\left.\delta\lambda\wedge\delta\log w\right|_{\gamma _k}=\sum _k\delta h_k
\wedge\delta\phi _k$ and the Hamiltonian flows provided by $h_k$ produce a
completely integrable system on Jacobian with co-\-ordinates $\{\phi _k\}$.
The symplectomorphisms of (\ref{dS}) can bring us to a {\it dual} system,
for example when $h_k$ become the time-variables themselves. Then
$dS$ (\ref{dS}) plays the role of the generating differential of the Whitham
hierarchy describing the flows in moduli space around a point, corresponding to
a finite-gap solution.

{\bf ${\cal T}$-function and prepotential}. The most complete information
about
the integrable system is given by the object $\log{\cal T} = \log{\cal T}_0
+ \log{\cal T}_{\theta}$ where the first part (a logariphm of a
quasiclassical $\tau $- function) restricted to the dependence
on moduli $\left.\log{\cal T}_0\right|_{moduli} \equiv {\cal F}$ is usually
called a {\it prepotential} and is defined by
\be\label{streq}
{\partial ^2{\cal F}\over \partial t_i\partial t_j} = T_{ij}(t)
\ee
The sense of parameters $t_k$ and the r.h.s. $T_{ij}(t)$ is
different for the different systems, but their geometrical meaning is
always related to certain "periods" $t_k = \oint _{A_k}dS$ or $t_{\alpha}
= \res _{P_{\alpha}}(\lambda ^{-\alpha}dS)$ and intersection form
$T_{ik} = \int _{\Sigma}d\omega _i\wedge d\omega _j$ on $\Sigma$.

The prepotential ${\cal F} = \log{\cal T}_0$ in general satisfies the
associativity equation having the form (for the matrices ${\cal F}_{ijk} =
{\partial ^3{\cal F}\over
\partial t_i\partial t_j\partial t_k} \equiv ({\cal F}_i)_{jk}$)
\be\label{assoc}
{\cal F}_i{\cal F}_j^{-1}{\cal F}_k = {\cal F}_k{\cal F}_j^{-1}{\cal F}_i
\ \ \ \ \ \forall i,j,k
\ee
The full generating (partition) function depends on the infinite amount of
variables -- related to all excitations of the effective theory (which should
include in particular gravitational dressing). It is defined by a
generalization of (\ref{streq}) -- usually called a {\it string} equation.
The string equation can be formulated
in terms of {\it quantization} of the symplectic structure (\ref{dS})
corresponding in general to a sort of quantization of the spectral curves:
where $\widehat{\cal L}$ and $\widehat Q \equiv \widehat{\log T}$ obey
$[\widehat{\cal L}, {\widehat Q}] = 1$.

{\bf "Toy string" solutions effective theories}. In this simplest example
the already known
{\it explicit} solutions exist for the spherical spectral curve $\Sigma $ where
only
the times related to the residues are valid. $T_{ij}(t)$ is a linear
function of the $t$-variables (giving rise to the prepotential
${\cal F} = {t_1^3\over 6} + \dots$) and the eigenvalues of two operators in
(\ref{dS}) are just polynomial functions. The topological correlators are
{\it numbers} and count the intersection indices on moduli space -- this
is an example of {\it topological} gravity.

The case of {\it physical} ($c<1$ or $pq$-) gravity is known much less
explicitely and correspond already to nontrivial spectral curves
$\Sigma _{g={(p-1)(q-1)\over 2}}$. This is however the case where the exact
form of the {\it duality} transformation -- relating the partition functions
in the dual points -- is known exactly, having the form of a Fourier transform
with the exponent $S = \int ^{\lambda}dS$.

{\bf The Seiberg-Witten effective theories}. The higher genus complex curves
arise also in $4d$ SUSY gauge theories where the nonperturbative exact solution
is formally defined as a map
\be
G,\tau ,h_k \rightarrow T_{ij}, \ a_i,\ a_i^D
\ee
$G$ is gauge group, $\tau$ -- the UV coupling
constant, $h_k={1\over k}\langle\Tr\Phi ^k\rangle$ -- the v.e.v.'s
of the Higgs field)
and an elegant description in terms
of $\Sigma _{g = \rank G}$ with $h_k$ parameterizing some of the
"hyperelliptic" moduli
of complex structures. The periods of meromorphic 1-form (\ref{dS})
$a_i = \oint_{A_i} dS$, $a_i^D = \oint_{B_i} dS$ determine the BPS massive
spectrum, $a_i^D=\frac{\partial{\cal F}}{\partial a^i}$ the prepotential
${\cal F}$ (giving the low-energy effective action) and, thus, the {\it set}
of low-energy coupling constants $T_{ij} = \frac{\partial^2{\cal F}}
{\partial a_i\partial a_j} =
\frac{\partial a^D_i}{\partial a_j}$.

The curves $\Sigma _{p = \rank G}$ are spectral curves of the nontrivial
finite-gap solutions of the periodic Toda-chain problem and its natural
deformations into Calogero-Moser and spin chains. The "period"-times $a_i$,
$a_i^D$ are related to the action integrals ($\oint pdq$) of the system.

{\bf String Duality}. The picture presented above should be actually
considered as a simple version of a generic nonperturbative effective
target-space formulation of {\it string} theory. String theory possesses
a huge amount of "hidden symmetries" which allow one sometimes to determine
the answer without a direct computation. The introduced objects have
a direct generalization for the whole string theory picture where at
the moment only some observations based on consistency requirements for the
relations among dual theories are
made. The difference of the presented above picture with generic conception of
string duality is that the above construction is formulated in strict
mathematical sense what still remains to be done for more "rich" string models.

A stringy generalization is straightforward and related first of all with the
prepotentials
arising in the study of realistic models related to the Calabi-Yau
compactifications. All the steps described above can be in principle repeated
leading finally to the integrable models based on the {\it higher-dimensional
complex} manifolds (instead of $1_{\bf C}$-dimensional $\Sigma$). Such
integrable systems (in spirit of Hitchin-\-Donagi-\-Markman) are not
investigated yet in such detail.

Another problem is that even for the simplest cases considered above the
complete picture has a lot of holes. In particular the exact form of the
generating function $\log{\cal T}$ is not yet known even for the
Seiberg-\-Witten effective theories.

In spite of all the problems it is easy to beleive that for all the theories
where it is possible to make any statement about the nonperturbative and
exact quantities there exists something more than a summation of a
perturbation theory. The main idea I tried to advocate above that this
could be the principle of {\it integrability}, which has been checked
already in several examples and based on general beleif that the realistic
theory should be a selfconsistent one and adjust automatically its
properties not to be ill-\-defined both at large and small distances.
It looks that an adequate language for the effective formulation of
nonperturbative field and string theories obeying such property can be
looked for among integrable systems.

\bigskip\bigskip
I am grateful to V.Fock, A.Gorsky, S.Kharchev, I.Krichever, A.Mironov,
A.Morozov, N.Nekrasov, A.Rosly, J.Schwarz, B.Voronov and A.Zabrodin
for the useful discussions. The work was partially supported by the
RFFI grant 96-02-16117 and INTAS grant 93-2058.

\end{document}